\documentclass[a4paper,twoside]{article}

\usepackage{epsfig}
\usepackage{subcaption}
\usepackage{calc}
\usepackage{amssymb}
\usepackage{amstext}
\usepackage{amsmath}
\usepackage{amsthm}
\usepackage{multicol}
\usepackage{pslatex}
\usepackage{apalike}
\usepackage{algorithm2e}
\usepackage[bottom]{footmisc}
\usepackage{SCITEPRESS}     
\usepackage{tikz}
\usepackage{xcolor}
\usetikzlibrary{arrows.meta,
                decorations.markings}
\begin{document}
\pagenumbering{arabic}

\title{Automation of Smart Homes with Multiple Rule Sources}

\author{\authorname{Hoffner Yigal, Kaufman Eran}
\affiliation{ Department of Software Engineering, Shenkar College, Ramat Gan, Israel }
\affiliation{\sup{2}Department of Software Engineering, Shenkar College, Israel }
\email{\{f\_author, s\_author\}@ips.xyz.edu, t\_author@dc.mu.edu}
}


\abstract{Using rules for home automation presents several challenges, especially when considering multiple stakeholders in addition to residents, such as homeowners, local authorities, energy suppliers, and system providers, who will wish to contribute rules to safeguard their interests. Managing rules from various sources requires a structured procedure, a relevant policy, and a designated authority to ensure authorized and correct contributions and address potential conflicts. In addition, the smart home rule language needs to express conditions and decisions at a high level of abstraction without specifying implementation details such as interfaces, access protocols, and room layout. Decoupling high-level decisions from these details supports the transferability and adaptability of rules to similar homes. This separation also has important implications for structuring the smart home system and the security architecture.\\
Our proposed approach and system implementation introduce a rule management process, a rule administrator, and a domain-specific rule language to address these challenges. In addition, the system provides a learning process that observes residents, detects behavior patterns, and derives rules which are then presented as recommendations to the system. 
}

\onecolumn \maketitle \normalsize \setcounter{footnote}{0} \vfill


\section{\uppercase{Introduction}}
\label{sec:introduction}

Using rules to automate the control of the home and its processes poses various challenges. Besides residents' rules, most homes are likely to have multiple stakeholders as additional sources of rules.  A stakeholder is an individual, group, or entity with an interest, involvement, or concern in the smart home, capable of affecting home-related decisions or being affected by their outcomes. Stakeholders, such as the homeowner, local council, the energy supplier, and the system provider, may also wish to contribute rules to protect and further their interests. For example, local authorities’ rules may reflect home regulations concerning such issues as pollution and permitted noise levels. An energy supplier may have requirements concerning the optimal time to use certain facilities. The homeowner may wish to keep the temperature in the house above 5 degrees, no matter what, to prevent pipes from freezing in the winter. The stakeholders must provide such rules, as they cannot be learned from observing the residents' behavior. 

The presence of multiple stakeholders as rule providers raises a management problem. Integrating rules from various sources requires a structured rule introduction procedure, a relevant policy and a designated authority, such as a Rule Administrator, to supervise it. This manager must ensure that the suggested rules originate from authorized entities and that any potential conflicts with other rules are addressed. This aspect of smart home management has yet to be discussed in the literature.

Another smart home problem concerns the rule language, which combines two central elements: conditions and decisions. In this approach, conditions refer to the home's state, its residents' activities, and contextual environmental factors. The corresponding actions are typically instructions to activate or deactivate specific home devices. A smart home rule language should be able to express automation preferences at a relatively high level of abstraction, enabling users to ignore the low-level technical implementation details and opening up the possibility of the same rules applying to similar houses. Decoupling the high-level decisions from the home layout details, the specific room allocation and the specific device used from the rules has considerable advantages. For example, if a rule sets the temperature at a certain level, this specifies a requirement but does not specify how to do it.  The requirement may be translated differently, depending on the house and the room's characteristics. For example, the required temperature may be obtained by turning the air conditioner, activating the ventilator, opening a window, or a combination thereof.

The high-level rules are more likely to be relevant to houses with a similar set-up. For example, rules defined for a family with kids are likely to suit other families who live in similar conditions. Such a scheme would make it easy for stakeholders to export their rules to different homes. Friends and neighbors would be able to recommend rule scripts to each other. In addition, the decoupling makes it possible to change the sensors or actuation devices without changing anything in the rules. 

Our approach addresses those challenges and introduces flexible system management methods and configurations. A Rule Administrator facilitates the integration of rules from multiple sources using a policy-driven rule introduction procedure. Our domain-specific rule language is designed to abstract from the specific details of the house configuration and the devices used, making it sufficiently general to be portable to comparable homes. Our system also supports a process that collects data from the system and uses it to learn the residents' behavior patterns and derive rules from them. In addition, the design of the rule language has important implications concerning the structuring of the smart home system and its security architecture and implementation. 

\subsection{Requirements of a Multi-Rules-Source Smart Home }

Several requirements can be derived from the above discussion concerning the smart home architecture:

\begin{itemize}
    \item A high-level rule language that abstracts away details such as room layouts, device configurations, device interfaces, and access details.
    
    \item A translation process from the abstract decision of the rules to home-specific actions, considering the home configuration, the available devices and their interfaces. Similarly, it must support the data collection, aggregation, and processing of the various sensors and residents’ input to provide the relevant information for decision-making.

    \item A structured and controlled policy-driven rule suggestion and incorporation process with a dedicated Rule Administrator who will ensure the related policy guides the process.

    \item Apart from the resident role, the system should accommodate various roles. The System Administrator configures, initializes, and oversees the computer system's overall operation, monitoring, and maintenance. The rule source or owner can be any stakeholder or an outcome of a learning process. The Rule Administrator role involves assessing proposed rules from different sources in the context of system policies, existing rules, and other relevant considerations.

    \item Hide the specific protocols of different device vendors behind a standard protocol. 
        
\end{itemize}

The structure of this paper is as follows. Section \ref{sec:overview} provides an overview of related systems. 
Section \ref{sec:architecture} introduces the architecture of our smart home systems. 
Section \ref{sec:managing} explains how endpoint devices are managed. 
Section \ref{sec:rules} shows how adding, deleting or editing rules are managed.  
Section \ref{sec:language} describes the rule language. 
Section \ref{sec:security} discusses the security issue which arises from this architecture.
Section \ref{sec:learning} describes how rules are derived from residents' behavior data. 
Section \ref{sec:implementation} outlines the implementation of our pilot system. 

\section{\uppercase{Overview of Related Systems}}
\label{sec:overview}

Many articles focus on monitoring and control technology by discussing the sensors and actuators connected through a microprocessor-based system. This approach involves integrating mobile phones with cloud networking via wireless communication. Users interact with cloud-based applications through their mobile devices, which serve as interfaces for monitoring and controlling various home appliances. Basic software facilities provide Some level of automation to allow the system to make decisions. The decision-making processes are, in most cases, embedded in the software. Commonly employed controllers for smart home automation systems include Arduino \cite{DBLP:conf/aqtr/GotaPFMV20}, Raspberry Pie \cite{DBLP:conf/nextcomp/SharmaAB22}, Microchip PIC \cite{DBLP:conf/iasam/ChekiredCTLBT21} and Nod MCU \cite{DBLP:journals/kbs/BahmanyarRM22}. These systems often provide the resident with a control panel to manage the house's devices.

In the architecture proposed by \cite {DBLP:journals/cm/XuWWSM16} for smart home systems, three levels are identified: the hardware layer, the controller layer, and the external service layer. The hardware layer comprises smart devices and sensors. The controller layer acts as a centralized management system, responsible for autonomously perceiving and analyzing user demands to manage the smart home. The external service layer integrates existing home service resources, delivering personalized services to the market. However, the specific structure of the controller and service layers, which handle most tasks, remains unclear. The Software-Defined Smart Home (SDSH) platform is built upon this architectural framework.

\cite {DBLP:journals/ijdsn/SinghRMKC19} propose a secure and efficient smart home architecture based on the cloud and blockchain technology. The architecture contains four components: smart home layer, blockchain network layer, cloud computing layer, and service layer. The Smart home layer comprises many IoT devices and other subsystems like security systems, control systems, home theaters, etc. In the Distributed blockchain layer, devices are managed by transactions and stored in the local blockchain. The Cloud layer provides services or applications. The service layer interacts with the service providers and smart home users. It focuses on application services offered by the cloud platform. Users use services or applications provided by smart home clouds, enterprise public clouds, or other third-party clouds.

\cite {DBLP:journals/access/MokhtariAZ19} present a cost-effective and scalable Smart Home Systems (SHSs) architecture using microcontrollers for remote automation and control of home appliances. The architecture comprises several layers: the Physical layer with various sensing technologies, the Fog-computing layer for basic data processing, the Network layer with gateways and communication protocols, the Cloud-computing layer for extensive processing and data communication, the Service layer offering operational and analytical data views, the Session layer facilitating data exchange between services, and the Application layer housing subscribed applications for data-driven services.

\cite {DBLP:journals/sensors/BurhanRKK18} propose a three-layered architecture. The Perception Layer identifies and gathers information from physical objects; the Network Layer serves as a bridge between the perception and application layers by transmitting collected information; and the Application Layer encompasses various applications utilizing IoT technology, such as smart homes, cities, health, and more, providing services to these applications. \cite {DBLP:journals/sensors/JaouhariPAB19} propose a similar architecture consisting of the Device, Gateway, and Application and Service layers.

All the layered architectures mentioned above describe the infrastructure supporting smart home systems without providing any information or help to structure the applications or services they are supposed to keep. As significant as the infrastructure is for smart home systems, the main challenge is the application's design. This is where our architecture and pilot smart home system differ: it offers a way to structure the system, not from an infrastructure point of view but from the application point of view.

There are quite a few articles about using rules for automating smart home decisions, but they ignore the possibility of multiple rule owners, the issues surrounding the management of the rules and how they interact with the rest of the system. The presence of rules and the need to interpret them and use their decisions as commands for activating devices considerably impact the system's structure. This point is missing from articles dealing with using rules in smart home systems. 

\section{\uppercase{The Smart Home Architecture}}
\label{sec:architecture}

The structure that underlies the architecture of our smart-home system is based on two ideas: 

\begin{itemize}
    \item \emph{Separation of concerns}: a design principle that manages complexity by partitioning the system into layers so that each layer is responsible for a different concern or set of concerns. Each layer has its decision-making process and is also responsible for aggregating the data and manipulating it to be used in the decision-making process. 
    \item \emph{Use of feedback}: Feedback refers to gathering data about a system's state or performance and then using it to make adjustments or corrections to the system. It helps maintain and improve the system's performance.
\end{itemize}

\begin{figure}[!ht]
  \centering
\includegraphics[width=8cm, height=4cm]{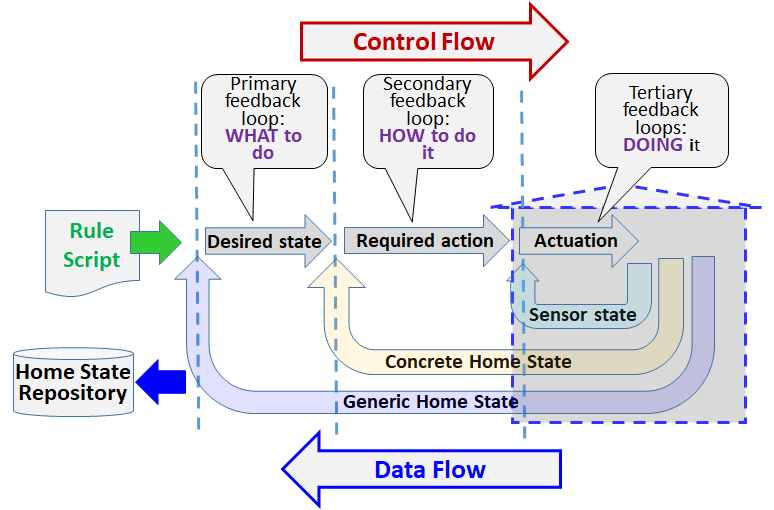}
\caption{The layered architecture of the smart-home system with three levels of feedback loops.}
  \label{fig:Layers}
 \end{figure}

The two ideas are combined to create a layered structure, where each layer consists of a feedback loop that encloses the next layer, as shown in Figure \ref{fig:Layers}. Each enclosing loop gets data from its inner layer and additional sensors for its decision-making and uses this data to make decisions for the enclosed loop. Each layer has its knowledge and responsibility. This results in the following feedback loops and subsequent layers:

\begin{itemize}
    \item \emph{The primary outer loop}: determines what should happen by using the rules to decide on the desired state of things.
    \item \emph{The middle secondary loop}: determines precisely what action should be taken, based on the decision made by the outer loop, and by which devices.
    \item \emph{The tertiary innermost loop}: activates the appropriate devices if so decreed by the secondary enclosing loop. 
\end{itemize}
The result is a three-layered architecture of the smart-home system, as shown in Figure \ref{fig:Layers}. 

There is two-way traffic between the layers: decisions flow in one direction and ultimately cause actions to be taken at end-point devices; data flows from the sensors and is aggregated and processed to create different views of the state of the home.

The three-layered architecture of our smart-home system uses two interdependent channels that flow in opposite directions through its layers: the data flow and the control flow channels.

\subsection{The Data Flow Channel}
\begin{figure}[ht]
  \centering
\includegraphics[width=\linewidth]{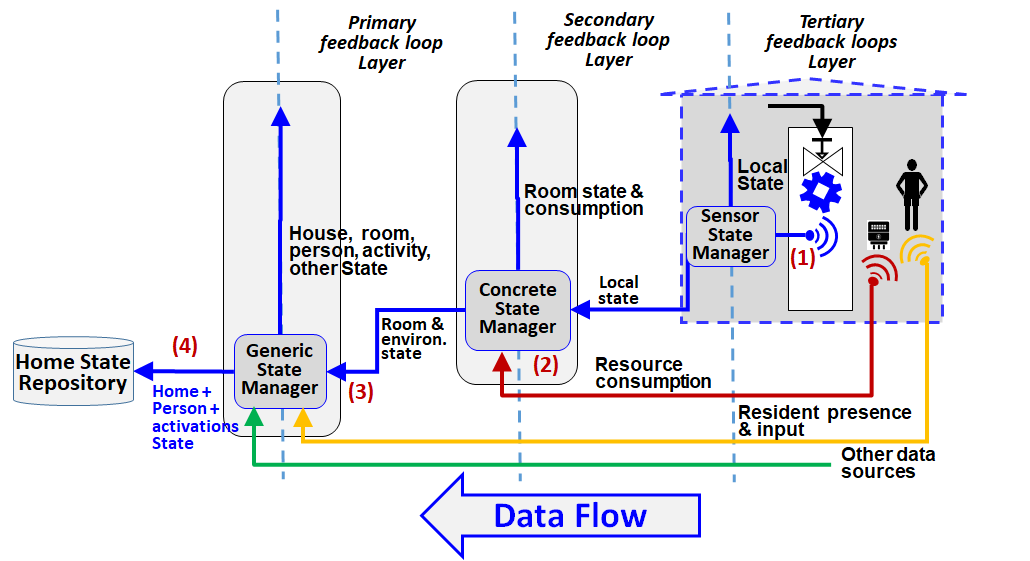}
\caption{The Data Flow channel, showing the stages that transform the data from the sensors into more comprehensive views of the system.}
  \label{fig:DataFlow}
 \end{figure}

The data flow channel (Figure \ref{fig:DataFlow}) transforms the data from the sensors to the appropriate level of abstraction so that it can be used by the decision-making process of the control flow channel. Thus, the data flow channel aggregates and analyzes data from sensors and other sources, such as residents’ audio input. It transforms individual data points into progressively more comprehensive views of the system, going from the status of a sensor to the condition of a room and, ultimately, to the overall state of the house and its current occupants’ activities.

The data flow channel comprises state-making managers, progressing from sensors to the Sensor State Manager, to the Concrete State Manager, and finally to the Generic State Manager. These managers play a central role in furnishing relevant data for the decision-making processes of the control flow, as shown in Figure \ref{fig:ControlFlow}:
\begin{enumerate}
    \item In the tertiary feedback loop, the sensors provide data for the local Sensor State Manager process to close the immediate feedback loop. A comparison of the set point of the loop with the current sensor value is used to decide whether to activate the connected actuator. At this level, the tertiary feedback loop does not know which room it is in or who is in the house; its sole responsibility is to reach and hold the set point dictated by its outer loop – the secondary layer. An example could be the room or the boiler’s water temperature or the ultraviolet level in the room. 
    \item The Concrete State Manager is responsible for aggregating the data from the tertiary layer Sensor State Manager and data about resource consumption from various meters. The process involves the Concrete Home Manager mapping devices to rooms, thereby putting together the state of rooms. At this level, the data passed to the Generic State Manager is, for example: Temperature in RoomX IS 25. 
    \item The Generic State Manager collects the data from the previous state managers together with data about the presence of residents in the house and some data about their activities. This is processed to create a high-level view of the state of the house, the persons present, which room they are in, and what activity they engage in. 
    \item The data from the Generic State Manager is fed to the Generic Home Manager and also sent to a Home State Repository for subsequent uses, which will be discussed later.
\end{enumerate}

\subsection{The Control Flow Channel}

\begin{figure}[ht]
  \centering
\includegraphics[width=\linewidth]{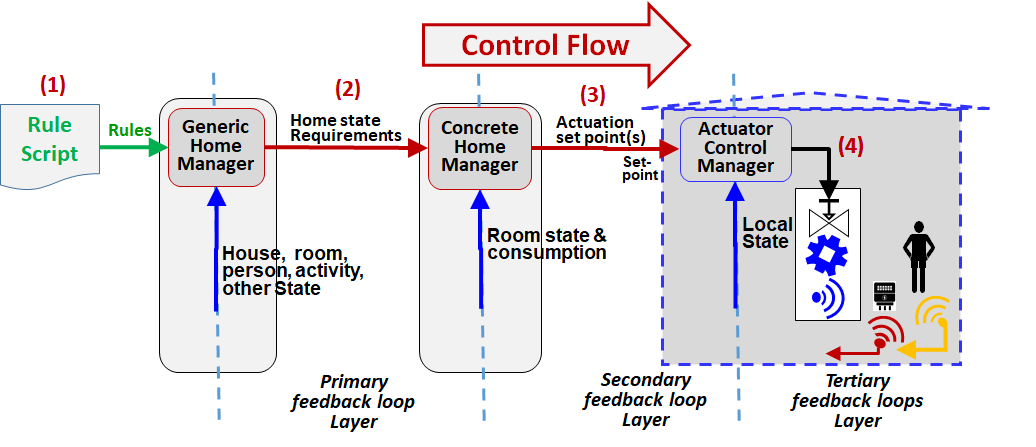}
\caption{The Control Flow channel and the translation between the layers going from the rules, through the decisions, to the activation of devices.}
  \label{fig:ControlFlow}
 \end{figure}

The control flow channel (Figure \ref{fig:ControlFlow}) comprises stages that make the appropriate decisions at each layer so that ultimately, the appropriate devices will be activated or deactivated based on the rule script and the state of the home. The Generic Home Manager consists of an Interpreter who can execute the rules in the rule script, thereby determining whether a home state change should be made. These decisions are passed to the Concrete Home Manager, ultimately becoming actions through the Actuator Control Manager. Those decisions flow from the high-level layer to the low-level ones, from ‘What to do’ to ‘How to do it’ and to ‘Doing it’, where each layer adds its relevant knowledge to the decision-making process.

The following are the stages of the control flow channel: 

\begin{enumerate}
  \item The primary feedback loop - the Generic Home Manager uses the rule script and the information from the Generic State Manager to determine if the state of the house needs changing.
  \item The state request is transferred to the Concrete Home Manager. It translates the request into commands to specific devices or devices.
  \item These commands can take one of two forms: as direct instructions to switch a device ON/OFF, for example, or as a set point or range to local feedback loops of the tertiary layer.
  \item The tertiary layer is a local feedback loop that is responsible for keeping the set point it received from the Concrete Home Manager.
\end{enumerate}

\subsection{{Combining the Data Flow and Control Flow channels}}

\begin{figure}[ht]
  \centering
\includegraphics[width=\linewidth]{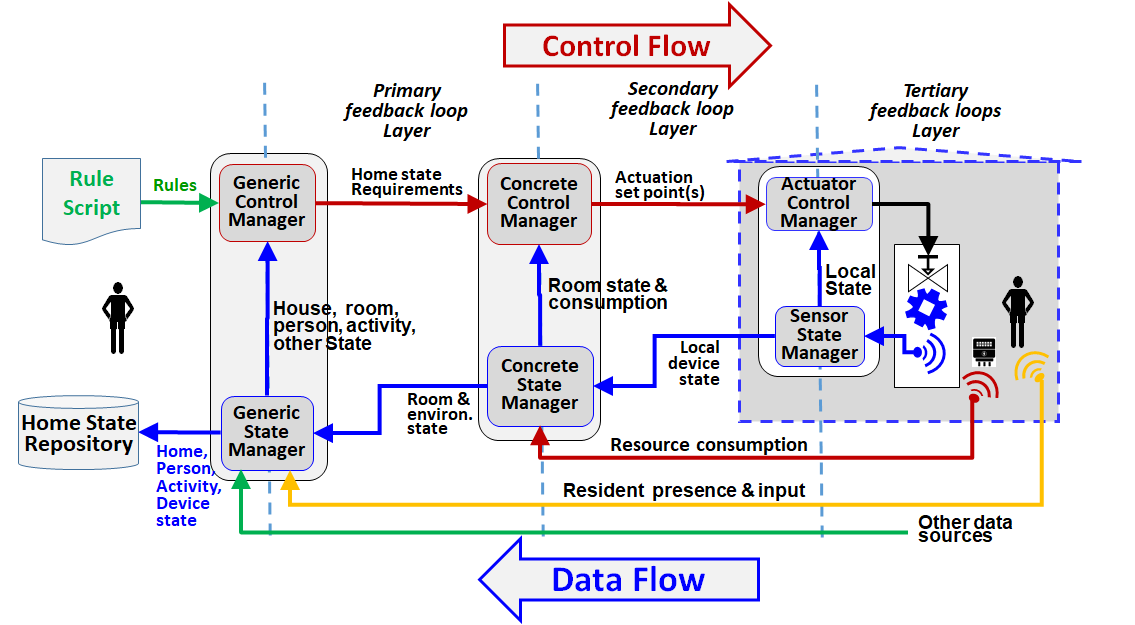}
\caption{The overall architecture of the smart home system shows both control flow and data flow channels.}
  \label{fig:Architecture}
 \end{figure}

By combining the data and control flows, we arrive at the smart home architecture shown in Figure \ref{fig:Architecture}. The data channel feeds the control channel with the information needed to make the decisions and the transformations in the Generic and Concrete Home Managers. Each engulfing feedback loop manages its inner feedback loop.

To explain how the architecture operates, we use an example of how rules work in the system. Given the following rule in the rule script that has been given to the Generic Home Manager to be executed by its Interpreter: 

       IF (Joe IN HOME AND SUMMER AND MORNING)
       THEN KEEP Joe ROOM\textunderscore TEMPERATURE KEEP BETWEEN 21 23

\begin{itemize}
    \item In the primary layer, data about whether Joe is in the house comes from sensors or as a result of input from Joe. This data is passed to the Generic State Manager. 
    \item Other sources of data, such as the date, time of day, and whether the day is a normal day, weekend day or holiday, are also passed to the Generic State Manager. 
    \item When the condition of the above rule is evaluated to TRUE, the command:
    
    KEEP Joe ROOM TEMPERATURE\textunderscore KEEP BETWEEN 21 23 is sent from the Generic State Manager to the Concrete State Manager. 
    \item In the secondary layer, the Concrete State Manager knows which room belongs to Joe and what devices are present. The request is translated to a set point to the cooling system to set the room temperature to 23. 
    \item The set point is sent to the specific Actuator Control Manager responsible for the cooling system. If the current set point of the air-conditioning system is already set to 22 degrees, no action is taken. 
    \item In the tertiary layer, if the cooling system is not self-regulating, a temperature sensor in the room will be used as part of a feedback loop. The current temperature will be compared to the set point given by the Concrete State Manager, and the difference will be used to decide whether to switch the cooling system. This will be the setup until conditions change and another rule is triggered.     
\end{itemize}
  
In a more complicated scenario, the Concrete State Manager knows that the window shutters are entirely open and decides to set the temperature to 23 and close the shutters a bit. 

\subsection{Management of End-Point Devices }
\label{sec:managing}

The Concrete Home Manager takes the high-level requirements from the Generic Home Manager and decides how to achieve them. Once the decision about which devices to activate is made, the Concrete Home Manager passes the request through the specific protocol to the selected device. 
Different devices will likely have their own protocol, as standards are slow to emerge or to be adopted in this domain. 

\begin{figure}[ht]
  \centering
\includegraphics[width=\linewidth, height=8cm]{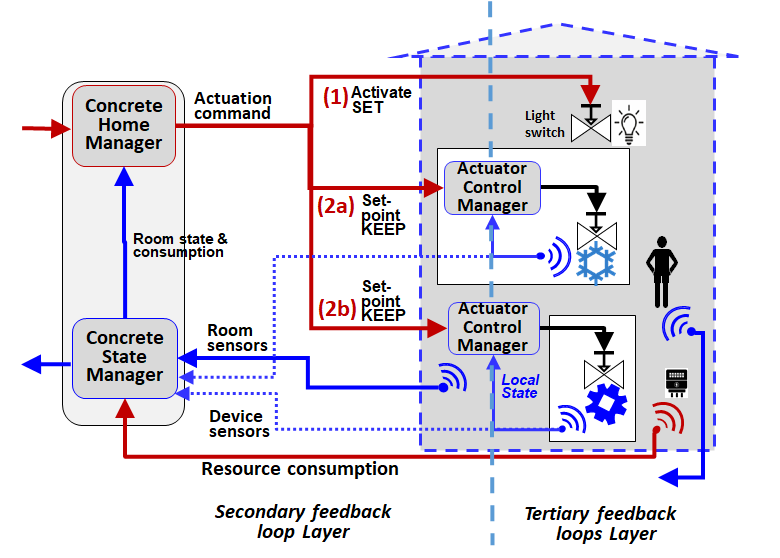}
\caption{Different management configurations of the actuator devices using the SET and KEEP commands.}
  \label{fig:TertiarySetUp}
 \end{figure}

In addition to the different protocols, actuator devices can have different management configurations, as shown in Figure \ref{fig:TertiarySetUp}, depending on whether there is a feedback loop that can continue to monitor the desired state:

\begin{enumerate}
  \item \emph{SET - By direct command}: Devices whose states are discrete can be set to a particular value by direct command. Examples are simple lights, ventilators, or laundry machines with on/off states. The command SET is used in this case. It expresses the discrete nature of the device it activates:
  
IF (Joe IN Home) THEN ‘SET Joe ROOM LIGHT ON’

IF (NIGHT) THEN ‘SET EXTERNAL\textunderscore DOORS CLOSE’

IF (Joe IN Home AND Joe ACTIVITY IS MUSIC) THEN ‘SET Joe ROOM MUSIC ON’

  \item \emph{KEEP - By desired set point}: Devices that control some variable that does not have a discrete value, such as temperature or humidity. Such devices require a feedback loop to reach and maintain the desired value provided by the set point. The command KEEP is used for this case. It expresses the requirement that a certain state is to be kept, using feedback, even if it drifts from the required value or range. There are two possible configurations in this category: devices whose feedback loop is already included as part of the device (Figure \ref{fig:TertiarySetUp} (1)) and devices whose feedback loop has to be constructed around the device (Figure \ref{fig:TertiarySetUp} (2a \& 2b)). Examples of the former are current air-conditioning systems with a feedback loop that maintains the required temperature given as a set-point input. Examples of the latter are humidity control or UV light control. 

IF (Joe IN HOME) THEN ‘KEEP Joe ROOM\textunderscore Temperature BETWEEN 23 21’
\end{enumerate}

\paragraph{Manual Overriding of Home State.}
There are situations where the residents wish to change the state of the home directly, possibly overriding the current settings. Such requests should enter the system through a special interface, such as a smartphone, connected to the Concrete Home Manager. This ensures that the user cannot access a specific device but only specify a desired state of some measured variables. This is further explained in Section \ref{sec:ourSecurity}. 
If the system includes a resident behavior learning component, then the changes should be monitored and recorded in the Home State Repository.  If such actions persist, they are likely to result in modifying the rules.

\section{\uppercase{The Rule Language}}
\label{sec:language}

The majority of rule-based systems adopt Trigger-Action programming paradigm in the form of "IF a trigger occurs WHILE conditions are true, THEN take some action" \cite{DBLP:journals/imwut/ZhangHMZL0U20,DBLP:journals/tochi/GhianiMPS17,DBLP:conf/chi/UrMHL14}. 

\begin{figure}[ht]
  \centering
\includegraphics[width=8cm, height=4cm]{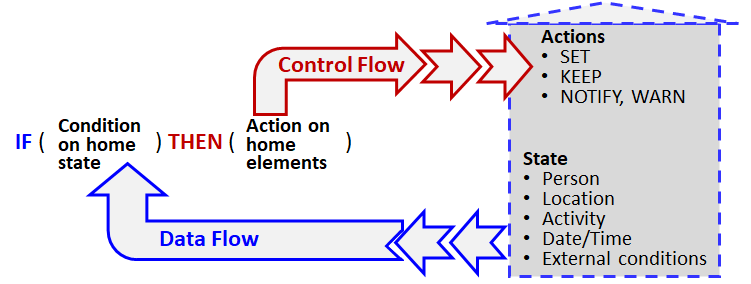}
\caption{The relationship between the data and control flow channels and the components of the rule.}
  \label{fig:FigLangRule}
 \end{figure}

Our rule language basic structure is a simpler conditional statement of the form: 
\emph {IF referral to Measured variables THEN referral to Control variables}.
This ties together the data-flow and the control-flow channels as shown in figure \ref{fig:FigLangRule}. The condition depends on the data supplied by the data-flow channel, and the decision sends a command to the next layer to act, thereby connecting to the control-flow channel.

The central idea of our language is to keep it sufficiently abstract so that the rules will be easily transferrable from one household to a similar one. While our rule language focuses on private homes, our approach can be applied to other domains, such as offices, hospitals and schools. While some keyword overlap is likely, each domain will also have its own special keywords. Schools will likely require keywords for classrooms, a staff room and an Office; Hospitals will need keywords for departments, patient rooms, receptions and laboratories.

The fundamental elements of our rule language include:

\begin{itemize}
 
  \item \emph{Domain Specific Keywords}: Reserved words with specific meanings are part of the language's syntax. The keywords in our rule language can be divided into several categories: Location, Role, Resident, Activity, Date/Time/Event and Action.

    There are two main sources of keywords:
    \begin{itemize}
        \item \emph{System default keywords}: These items are expected to be present in almost every house and are therefore provided as default words. This is likely to make the rule scripts more transferable and readable.
        \item \emph{User-defined keywords}: Stakeholders can add keywords if they fit the particular circumstances of their homes.
    \end{itemize}
     
    The keywords can be divided into several categories:
    
    \begin{itemize}    
        \item \emph{Location}: A place inside or outside the house.  Private home defaults can be Home, Kitchen, LivingRoom, BedRoom etc. and special terms like AllRooms : Home, Kitchen, Living room.
            Other user-specific locations in the system can be added by a user to the application by clicking 'add location' and adding the chosen name of the room. \\
            Rule example: \emph {IF (Joe IN Bathroom) AT NOON}
    
        \item \emph{Role}: A person related to the house with a set of responsibilities, accountabilities, liabilities, and permissions granted through the security setup (see section \ref{sec:security}).
            These could include defaults such as Resident, Homeowner, CleaningPerson, MaintenancePerson and EnergySupplier.
            A user can add other users to the system by clicking 'add role' with the name, phone number and other identification.\\
            Rule example: \emph {IF (CleaningPerson IN Bathroom) THEN SET LightSET IN Bathroom} ON. 
        
        \item \emph{Resident}: These are the specific names of the house's residents. Examples: Joe, Ruth, Spencer
    
        \item \emph{Activity}: These activities are common in houses and related to a person or persons. 
                Activities can be Sleeping, Resting, Exercising, Working, etc.  
                User-specific activities may be added. \\
                Rule example: \emph {IF Anyone IS Sleeping THEN SET AllVolume Below 25}
    
        \item \emph{Date/Time/Event}: These are indications of time or events that make a difference in the residents' behavior. \\
        System default:AM, PM, Morning, Afternoon, Evening, Night, Holiday, Xmas, Easter, Weekend, Today, Tomorrow, Minute, Hour, Day, Week, Month, Year, Always
    
        \item \emph{Action}: These commands result from decisions made by the THEN part of a conditional statement. 
            System default actions: SET, KEEP, ON, OFF, CLOSE, OPEN, NOTIFY, WARN.\\ 
            Warnings and notifications are special cases of action.  This is relevant when the residents should be made aware of the state of the house.  Security-related notifications are one example of such warnings. \\
            For example, the rule: \emph {IF Anyone IN Home AND AllTenants NOT IN Home THEN WARN Joe}.
            The first part of the conditional statement could be, for example, triggered by a motion detector and the second part by locating the residents' phones.
    \end{itemize}
    \item \emph{Variables}: Variables are used to store and represent data in a system. They have names and data types, and their values can be changed during the system's lifetime. There are two types of variables in our system:
            \begin{itemize}
                \item \emph{Measured variables}: These variables depend on the specific things that are of interest and can be measured in a home. Ultimately, each item is linked to one or more sensors or input devices, but the rule language does not have to be aware of the sensors or input devices, only the data provided by the data flow channel. By agreed practice, such variables have the postfix VAL. \\
                Examples: TemperatureVAL, HumidityVAL, UVLevelVAL, WindowVAL, SmokeVAL.\\
                Rule example: \emph {IF TemperatureVAL IN Kitechen ABOVE 25 THEN}
            
                \item \emph{Controlled variables}: These variables depend on the things that can be actuated in a home either by a direct command or set-point input to a feedback loop. By agreed practice, such variables have the postfix KEEP or SET.\\
                Examples: TemperatureKEEP, HumidityKEEP, UVLevelKEEP, DoorLockSET, LightSET, LaundryMachineSET.
            \end{itemize}

    \item \emph{Operators}: Operators perform operations on variables and values. 
        \begin{itemize}
            \item \emph{Relational operators}: EQUAL, ABOVE, BELOW.
            \item \emph{Logical operators}: AND, OR, NOT.
        \end{itemize}

    \item \emph{Control Structures}: Control structures determine the flow of execution shown in figure \ref{fig:FigLangRule}.\\
    Rule example: \emph {IF (referral to Measured variables) THEN (referral to Control variables)}

 \end{itemize}
 
\section{\uppercase{The Rule Management}}
\label{sec:rules}

\begin{figure}[ht]
  \centering
\includegraphics[width=\linewidth]{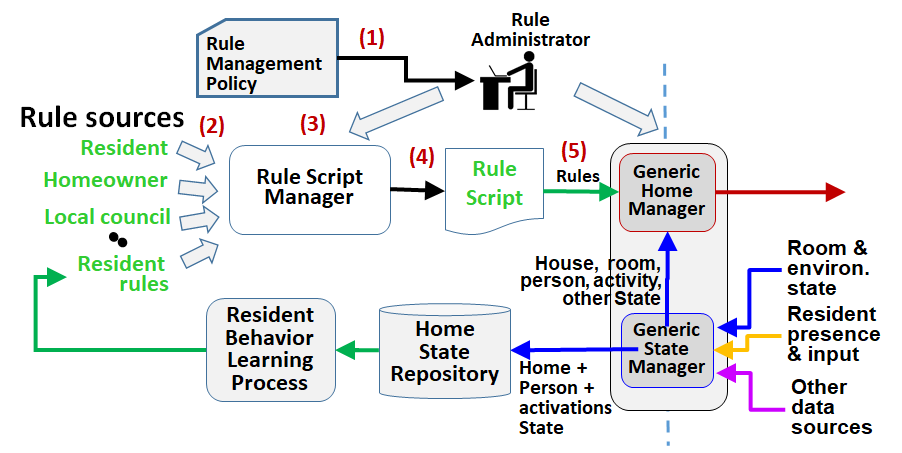}
\caption{The Rule Management Policy determines how to deal with any rule proposals from the various stakeholders.}
  \label{fig:RuleMgt}
 \end{figure}

The Rule Script Manager shown in Figure \ref{fig:RuleMgt} is concerned with adding, editing or deleting a rule in the Rule Script that is ultimately executed by an Interpreter in the Generic Home Manager. This process has a Rule Management Policy and a Rule Administrator associated with it that determine the following things:

\begin{itemize}
    \item \emph{Permissions}: Specifies which users are authorized to view the home state variables and request a change of control variables as written in the rule.
    
    \item \emph{Conflict resolution procedure}: The procedure for resolving conflicts \cite {DBLP:journals/percom/CarreiraRS14} can include the following: dropping the rule, sending a warning back to the source, or referring the resolution to a higher hierarchy. This will be further explained in Section \ref{sec:RBAC}.
    
\item \emph{Rule script update}: The point at which it is possible to replace the rule script of a working system with an updated one.
\end{itemize}

Figure \ref{fig:RuleMgt} is an example of the rule management process when a rule recommender tries to add a new rule to the rule script:

\begin{enumerate}
  \item The administrator gives the rule management policy to the Rule Administrator. 
  \item One of the authorized stakeholders suggests incorporating a new rule into the rule script.
  \item The Rule Administrator uses the Rule Script Manager to check the rule from several points of view:
  
\begin{itemize}
    \item The rule has the correct syntax.
    \item The stakeholder is authorized to read from and write to the designated devices according to the policy.
    \item The new rule does not conflict with an existing rule in the currently executing script; if it does, the Rule Administrator resolves it according to the conflict resolution given in the policy.
\end{itemize}
  
  \item The rule is added to the home rule script.
  \item At a quiescent point in the running of the Generic Home Manager Interpreter, the new script replaces the old one, according to the rule update policy.
\end{enumerate}

The Rule Script Manager shown in Figure \ref{fig:RuleMgt} can be implemented differently, depending on the system's goal. The two extreme points on the spectrum of possible designs are:

\begin{itemize}
    \item \emph{Complete human control}: This would require an interface that presents the new rule to the Rule Administrator, who has complete control over the process. 
    \item \emph{Maximum automation}: This would require embedding the policy and the process in software.  Nevertheless, this solution must still provide a minimal interface to the Rule Administrator in case of problems that the automation cannot resolve.
\end{itemize}
If the implementation is automated, then a security mechanism should be in place. This issue is elaborated upon in section \ref{sec:security}

\section{\uppercase{The Learning Process}}
\label{sec:learning}

\begin{figure}[ht]
  \centering
\includegraphics[width=8cm, height=4cm]{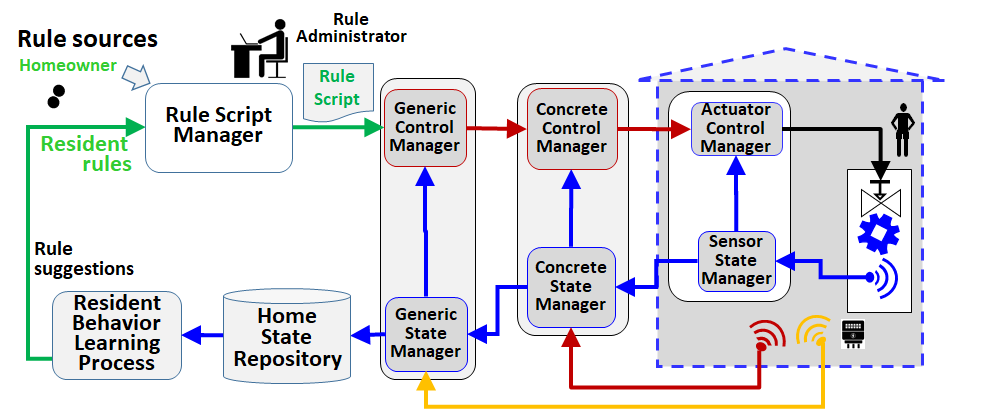}
\caption{Closing the loop with data analysis such as the learning process.}
  \label{fig:LearningProcess}
 \end{figure}

The purpose of the learning process is to derive rules in the language defined in section \ref{sec:language}, based on the residents' activities in the house. The learning process can act as a source of rules for the rule management process, as shown in figures \ref{fig:RuleMgt} and \ref{fig:LearningProcess}.

\subsection{Related Work}
Given a sequence of user behaviors, a sequential recommendation system aims to recommend items to users by modeling sequential dependencies of the user behavior \cite{DBLP:conf/ijcai/WangHWCSO19}. For example, FPMC \cite{DBLP:conf/www/RendleFS10} combines first-order Markov chains (MCs) and Matrix Factorization \cite{KorenBV09,ZhanHP022} to model both sequential behaviors and general interests of users. 
Besides the first-order MCs, higher-order MCs \cite{ZhanHP022} have been studied to consider previous items in a sequence.

In recent years, deep neural networks such as Recurrent Neural Networks (RNN) \cite{ChoMGBBSB14}, Convolutional Neural Networks (CNN), and Transformers have been adopted in sequential recommendation to address the complicated non-linear patterns in user behaviors.
For instance, GRU4Rec \cite{HidasiKBT15} used Gated Recurrent Unit (GRU) to model sequential patterns for session-based recommendation.

Caser \cite{TangW18} employed CNN to capture sequential patterns from both the time-axis and feature-axis of sequences.
SAS-Rec \cite{RenLLZWDW20} and BERT4Rec \cite{SunLWPLOJ19} utilized unidirectional Transformers and bidirectional Transformers, respectively, to capture sequential patterns in sequences while considering the importance of correlations between behaviors.

Context-aware recommendation aims to capture user preferences by considering contextual attributes such as time and temperature as well as interaction information between users and
items \cite{KulkarniR20}.

These include methods such as Factorization Machines (FMs) \cite{RendleGFS11,XinCHWDJ19}, which model all correlations between variables to capture the significant correlations between contexts.

Deep neural attention mechanisms \cite{XiaoY0ZWC17} to model higher-order interactions and generate more meaningful representations of contexts. Context-aware sequential recommendation, which considers sequential patterns as well as contextual information, such as CA-RNN \cite{LiuWWLW16} which employs a context-specific transition matrix to represent contextual information and adopts RNN to capture sequential patterns in history, Analogously, SIAR \cite{RakkappanR19} utilizes stacked RNN to consider the temporal dynamics of both contexts and actions. 

\subsection{Our Learning Process Implementation}

In our implementation, we used Markovian belief nets to capture context-aware reasoning. 
The house has two main components: measured variables and controlled variables.
These two components are intricately connected through the user's activity and provide formal reasoning of the form:
'IF $SENSOR_1$ EQUAL 5 and $SENSOR_2$ ABOVE 25 THEN SET $ACTUATOR_1$ ON and KEEP $ACTUATOR_2$ ABOVE 25 '.

\subsection{Reasoning Premise}
Our premise is that the house rules follow the argument that sensor readings always cause device actuation and are not the effect. For example, the presence of a person in a room is the reason for the light being turned on, not the other way around. The exception is when sensors measure the state of the actuator for control reasons, in which case those sensors are considered part of the actuator. For example:
'IF HumidityVAL above 15 THEN SET IrrigationSET ON' and never 'IF irrigation on THEN humidity is above 15'.
Time is always the cause so 'IF 2AM THEN SET Laundry ON' and never: 'IF Laundry ON THEN 2 AM'.

The reasoning premise is encoded into the Markovian network structure.
The actuator's state is being reported along with other residual sensed parameters such as heat, humidity, presence etc.
By using reinforcement learning, the connections that co-occur at a certain time were strengthened while others were weakened.
For example, every time Joe is in a room and the room temperature is above a certain degree, Joe turns on the AC at a preferred level.
In this example, the activation of the AC co-occur with the relevant measured variables such as humidity and temperature with high probability, while other measured variables such as light or pressure act as noise for this actuator's state.

\subsection{Markovian networks background}
A Markovian network for a set of variables $X = \{X_1, . . . , X_n\}$ is the pair $(G, \theta)$ over $X$, where $G$ is a directed acyclic graph with $n$ nodes, one for each variable in $X$, called the network structure, and $\theta$ is a set of conditional probability functions, called the network parametrization.

Each variable only directly depends on its predecessor in the sequence.
For each variable, its parameter is the set of conditional probability functions of the variable given each value of its parents in the graph.
According to the type of random variable, three types of Bayesian networks are distinguished: discrete if the variables are discrete, continuous if the variables are continuous, and hybrid if there are both discrete and continuous variables in the network.

In our model, a hybrid network is considered.
The graph encodes a set of conditional independence assumptions called local independence:
Each variable $X_i$ is conditionally independent on its non-descendants given its parents in the graph.
An important result is the chain rule, which states that:
\begin{equation}
    P(X_1,\ldots,X_n)=\prod P(X_i | Pa(X_i))
\end{equation}
where $Pa(X_i)$ are the parents of $X_i$ in the graph.

This method states that given a data set of instantiations of the variables in a discrete Bayesian network, each probability of the parameters is estimated with the formula:
\begin{equation}
    \theta_{x_i|pa(x_i))}= \frac{D(x_i,pa(x_i)}{D(pa(x_i))}
\end{equation}
where $D(x_i, pa(x_i))$ is the number of cases in the dataset that contain the instantiation $x_i$ of $X_i$ and $pa(x_i)$ of its parents $Pa(Xi)$; and $D(pa(x_i))$ is the number of cases in the data set that contain the instantiation $pa(x_i)$ of $Pa(Xi)$.
Once the Bayesian network has been designed, i.e., the structure and parameters have been specified, the objective is to obtain information by answering questions about the variables and their probabilities. The techniques used are known in general as inference. 
Given a Bayesian network for a set of random variables $X = \{X1, . . . , Xn\}$ these inferences are:
\begin{itemize}
    \item Probability of evidence: $P(Xi = xi) \quad \forall  x_i \in \{0,1\},\quad i \in \{1,\ldots,n\}$.
    \item Posterior marginals: Let $X_j = x_j$, be evidence of the Bayesian network, posterior
marginals are $P(X_i=x_i|X_j=x_j) \quad \forall x_i \in \{0,1\}, \quad \forall i \in \{1,\ldots,j - 1,j + 1,\ldots,n\}$.
\item Most probable explanation: Let $X_n = x_n$, be evidence of the Bayesian network, most probable explanation are 
$x_i \in \{0,1\}, \forall i \in  \{1,\ldots,n-1\}$ 
such that $P(X_1 = x_1, \ldots,X_n = x_n )$ is maximum.
\end{itemize}

\subsection{From a Network to a Rule Language}

\begin{figure}
  \centering
\includegraphics[width=0.5\textwidth]{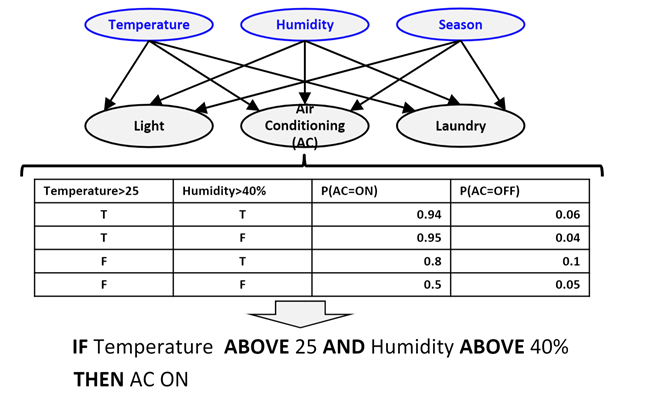}
\caption{An example of network structure and reasoning.}
  \label{fig:learn}
 \end{figure}

In our case, we used the posterior marginal between sensors and actuators to create new rules.
We used a very simplified architecture in our implementation, where the sensors are fully connected and directed to all actuators (see figure \ref{fig:learn}.
Other structures can be considered involving latent variables, which may represent certain learned latent states.

When a certain threshold of the joint probability is reached, a recommendation connecting these parameters appears.
The system will update these probabilities (online learning) whenever new data becomes available, ensuring that the system's recommendations evolve over time and adapt
to changing conditions.

If the user regards this rule as being unwanted, then the threshold is raised so the rule does not appear again in the recommendations system. This indicates that the consistency for this activity needs to be stronger in order for it to be reaffirmed.

\section{\uppercase{Security and Privacy}}
\label{sec:security}

One of the innovative principles of our design is the way access control is used to determine whether a specific user can or cannot view some measured variables or manage the state of some controlled variables.
In our approach, there is no direct reference to concrete devices. Instead, the user specifies the desired state of some controlled variable, which the Concrete Control Manager translates to specific device activation. The result is that access control is done based on state requests rather than direct access requests to devices. Thus, there is no need to specify device access permissions for access control. For example, if a user wants to change the temperature of a room that does not belong to her, the access control process would refuse to authorize it.



\subsection{Related work} In order to tackle the security attacks that threaten the privacy and safety of the smart home, several architectures have been proposed.
\cite{DBLP:conf/milcom/ParkDISSK16} presents a traffic monitoring and inspection solution called IoTGuard. 
All traffic is routed to the IoT Controller and linked to each IoT device with the IoT Watchdog in order to target and monitor particular IoT devices using device-specific IoT protocols.
The Intrusion Detection System (IDS) framework of \cite{DBLP:conf/saso/PachecoH16}, is based on Anomaly behavior analysis and tries to provide security by monitoring the traffic that runs through the main gateway.

The work by \cite{DBLP:conf/iciot/RaffertyIA0HH18} relies on Agent-based modeling, where agents inside the smart home environment make observations and implement the intended behavior. This model requires minimal user engagement and is focused on threat detection. 

Habibi et al., \cite{DBLP:journals/iotj/HabibiMMB17} proposed a whitelist-based intrusion detection technique specific to IoT devices. The proposal aims to prevent IoT devices from getting entangled in botnet activities, so it blocks DNS lookups to malicious sites at the gateway level. 

Serror et al. \cite{DBLP:conf/IEEEares/SerrorHHSW18} follow a rule-based approach, where every IoT device is allowed a specific behavior, namely the specific set of allowed connections, in order to fulfill its intended functionality. The gateway enforces these rules with traffic filtering and anomaly detection techniques.

\subsection{Our Security Approach}
\label{sec:ourSecurity}

As the smart home is a distributed system, each module should employ the minimum trust policy.
In other words, communication should be encrypted, users should be authenticated, and actions authorized using a suitable policy. This follows along the lines of the 3 A's of the golden standard for access control: authenticate, authorize, and audit.


As mentioned in section \ref{sec:rules} - a rule management policy is made out of the following:
\begin{itemize}
    \item \emph{Permissions}
    \item \emph{Conflict resolution}
    \item \emph{Rule  update}
\end{itemize}
 
The Concrete State Manager serves as a gateway connecting to the outside world and abstracting the local home devices and their proprietary protocols, which reside inside a local net.
Permissions are related to users and states through the \emph{Access Control Lists} (ACL).
The ACL is made out of state users and their related actions for example:

\begin{center}
\begin{tabular}{||c c c c|| }
 \hline
 State & User & Action & Value \\  
 \hline\hline
 Temperature & resident & Read & All \\ 
 \hline
 Temperature & resident & SET & ABOVE 5 \\
 \hline
 Lights & resident & Read \& SET & Any \\
 \hline
 Temperature & Owner & Read \& SET & Any \\
 \hline
Lights & Owner & Read \& SET & None \\ \hline
\end{tabular}
\end{center}

The security modules are therefore made up of two main components: the \emph{Authentication Service} (AS) and the \emph{Access Control Service} (ACS).

\paragraph{The authentication process.} 
Authentication can be done via different schemes, such as passwords, second-factor authentication, etc. This service can also be a third-party service.
The AS grants a ticket signed by the AS private key to the client, which serves as a client certificate for all the following actions. The AS signs the user IP (and/or associated domains) and user name (as it appears in the ACL) with a time stamp, expiration date, and a nonce.

\paragraph{The authorization process.} 
The client sends a request to the ACS using the certificate granted by the AS which verifies his identity.

The ACS verifies the AS signature of the client certificate and uses the ACL to grant a ticket for the requested state's read or modify according to the policy with the user's credentials signed by its own private key.

The signature holds the user IP (and/or associated domains) and user name (as it appears in the ACL), the desired state name (as it appears in the ACL) and the requested action (read or set).

Stakeholders have two main ways to access home devices:
\begin{itemize}
    \item\emph{Overriding current home state}: Using remote access with a tablet or smartphone in order to change the state of one of the home-controlled variables, such as temperature or lighting, as a 'one-off' request. The request goes to the Concrete Home Manager after getting a ticket from the authorization process. The Concrete Home Manager guard validates their ticket using the ACS public key. It then translates the requested state change to the activation of the necessary device and activates the selected device through its proprietary API.
    
    \item\emph{Adding a rule to the rule script}: Requesting a conditional change to one of the states of the house. This is done through the rule-management process.
    
When stakeholders want to add a rule, they have to go through the rule management process described in Section \ref{sec:rules}.
The rule management process guard checks the granted ticket for access to all READ and SET state requests involved in this rule. If all is well, it adds the rule to the rule's list with the user's name as the rule source. This completes stage 3 of the rule management process.
\end{itemize}

When the rule's condition is satisfied at run time, the Generic Home Manager turns to the Concrete Control Manager with the request to change a house state on behalf of the user. The Generic Home Manager signs the request with its own private key. Since it has already checked the user's permissions in the Rule Script Manager, there is no need for further permission checks.

\subsection{Using User Priority for Rule Conflict Resolution}
\label{sec:RBAC}


The diversity of rule owners raises the possibility of clashes among rules suggested by different rule owners.
One way of resolving this issue is by assigning priorities to roles and using the priority to decide in favor of the higher priority. Such a scheme affords automation of the resolution process. 

One possible way to implement this is to create a hierarchy of rule owners with the Rule Administrator as the tree's root. Priority is given to users according to their level in the tree.
This exploits the hierarchical order to resolve conflicts between rules. In cases of conflicts \cite {DBLP:journals/corr/abs-2310-04447}, priority is given to the parent over the children. The parent's rules always supersede the rules given by their children.

The Rule Administrator can create the tree, or alternatively, each added rule owner can add rule owners as its children, assigning priorities automatically. In most cases, where the sources of the conflicting rules have different priorities, the rule with the higher priority source will supersede the lower one. Problems arise when the sources of the contradictory rules have the same priority. In these cases, the resolution of the conflict is referred to the closest shared parent.


\paragraph{Examples.}
Here are some examples of house rules, hierarchies, and desired conflict resolution and how they can be implemented:  
\begin{itemize}
    \item The resident has permission to change the house temperature. However, in order to prevent the water pipes from freezing at winter time, the house temperature can not decrease below a certain level.
    In this case administrator sets up the homeowner as a user and as a Rule Administrator.
    The homeowner adds the resident as a user, in which he restricts his access to the house temperature to be above 5 degrees.

\item The electricity provider company and the municipality can be users with recommend-only permission, in which case power-saving tips like 'IF AT 2 A.M, THEN SET LaundryVal ON' will be treated as recommendations only.
They can always move higher in the hierarchy if the Rule Administrator wishes to make their rules obligatory.
They can restrict the resident's ability to change the music volume above a certain level at certain hours while still being restricted to being unable to read some of the home states for privacy reasons.
\end{itemize}

\section{\uppercase{The Smart Home System Implementation}}
\label{sec:implementation}

Our architecture was validated through a proof-of-concept implementation of a smart home system. 
The server is based on NodeJs with React and the learning microservice was written in Python with a Restful API.
The Concrete Control Manager and Concrete State Manager reside inside a local network located at home. The Generic Control Manager and Generic State Manager reside remotely in the cloud.
The data collected by the different state managers is stored in the Home Data Repository, and the rule script is stored in the Rule Script Repository. Both repositories are implemented using MongoDB.

We modeled a smart home with the following devices:
\begin{itemize}
    \item AC from Sensibo using its HTTP-based interface.
    \item Tuya washing machine with its HTTP-based interface.
    \item Tuya smart plug for the illumination system.
    \item A proprietary irrigation system for pots and plants with humidity and temperature sensors and water pipe actuators connected to an ESP32 micro-controller with an MQTT protocol interface.
\end{itemize}

The system implementation code can be found here \footnote{Client repository: https://github.com/harelfogel/SmartByte-Alpha.git
Server(general home manager) repository: https://github.com/harelfogel/SmartByte-Alpha.git
Concrete Home Manager repository: https://github.com/harelfogel/Alpha-Interpreter.git
Learning Process repository: https://github.com/YovelElad/SmartByte-Training-Data.git}.


\newpage

\bibliographystyle{apalike}
{\small
\bibliography{example}}

\begin{thebibliography}{}

\bibitem[Bahmanyar et~al., 2022]{DBLP:journals/kbs/BahmanyarRM22}
Bahmanyar, D., Razmjooy, N., and Mirjalili, S. (2022).
\newblock Multi-objective scheduling of iot-enabled smart homes for energy management based on arithmetic optimization algorithm: {A} node-red and nodemcu module-based technique.
\newblock {\em Knowl. Based Syst.}, 247:108762.

\bibitem[Burhan et~al., 2018]{DBLP:journals/sensors/BurhanRKK18}
Burhan, M., Rehman, R.~A., Khan, B., and Kim, B. (2018).
\newblock Iot elements, layered architectures and security issues: {A} comprehensive survey.
\newblock {\em Sensors}, 18(9):2796.

\bibitem[Carreira et~al., 2014]{DBLP:journals/percom/CarreiraRS14}
Carreira, P., Resendes, S., and Santos, A.~C. (2014).
\newblock Towards automatic conflict detection in home and building automation systems.
\newblock {\em Pervasive Mob. Comput.}, 12:37--57.

\bibitem[Chekired et~al., 2021]{DBLP:conf/iasam/ChekiredCTLBT21}
Chekired, F., Canale, L., Tadjer, S., Louni, A., Bouroussis, C.~A., and Tilmatine, A. (2021).
\newblock Low cost house automation system based on arduino microcontroller.
\newblock In {\em {IEEE} Industry Applications Society Annual Meeting, {IAS} 2021, Vancouver, BC, Canada, October 10-14, 2021}, pages 1--6. {IEEE}.

\bibitem[Cho et~al., 2014]{ChoMGBBSB14}
Cho, K., van Merrienboer, B., G{\"{u}}l{\c{c}}ehre, {\c{C}}., Bahdanau, D., Bougares, F., Schwenk, H., and Bengio, Y. (2014).
\newblock Learning phrase representations using {RNN} encoder-decoder for statistical machine translation.
\newblock In Moschitti, A., Pang, B., and Daelemans, W., editors, {\em Proceedings of the 2014 Conference on Empirical Methods in Natural Language Processing, {EMNLP} 2014, October 25-29, 2014, Doha, Qatar, {A} meeting of SIGDAT, a Special Interest Group of the {ACL}}, pages 1724--1734. {ACL}.

\bibitem[Ghiani et~al., 2017]{DBLP:journals/tochi/GhianiMPS17}
Ghiani, G., Manca, M., Patern{\`{o}}, F., and Santoro, C. (2017).
\newblock Personalization of context-dependent applications through trigger-action rules.
\newblock {\em {ACM} Trans. Comput. Hum. Interact.}, 24(2):14:1--14:33.

\bibitem[Gota et~al., 2020]{DBLP:conf/aqtr/GotaPFMV20}
Gota, D., Puscasiu, A., Fanca, A., Miclea, L., and Valean, H. (2020).
\newblock Smart home automation system using arduino microcontrollers.
\newblock In {\em {IEEE} International Conference on Automation, Quality and Testing, Robotics, {AQTR} 2020, Cluj-Napoca, Romania, May 21-23, 2020}, pages 1--7. {IEEE}.

\bibitem[Habibi et~al., 2017]{DBLP:journals/iotj/HabibiMMB17}
Habibi, J., Midi, D., Mudgerikar, A., and Bertino, E. (2017).
\newblock Heimdall: Mitigating the internet of insecure things.
\newblock {\em {IEEE} Internet Things J.}, 4(4):968--978.

\bibitem[Hidasi et~al., 2016]{HidasiKBT15}
Hidasi, B., Karatzoglou, A., Baltrunas, L., and Tikk, D. (2016).
\newblock Session-based recommendations with recurrent neural networks.
\newblock In Bengio, Y. and LeCun, Y., editors, {\em 4th International Conference on Learning Representations, {ICLR} 2016, San Juan, Puerto Rico, May 2-4, 2016, Conference Track Proceedings}.

\bibitem[Huang et~al., 2023]{DBLP:journals/corr/abs-2310-04447}
Huang, B., Chaki, D., Bouguettaya, A., and Lam, K. (2023).
\newblock A survey on conflict detection in iot-based smart homes.
\newblock {\em CoRR}, abs/2310.04447.

\bibitem[Jaouhari et~al., 2019]{DBLP:journals/sensors/JaouhariPAB19}
Jaouhari, S.~E., Palacios{-}Garc{\'{\i}}a, E.~J., Anvari{-}Moghaddam, A., and Bouabdallah, A. (2019).
\newblock Integrated management of energy, wellbeing and health in the next generation of smart homes.
\newblock {\em Sensors}, 19(3):481.

\bibitem[Koren et~al., 2009]{KorenBV09}
Koren, Y., Bell, R.~M., and Volinsky, C. (2009).
\newblock Matrix factorization techniques for recommender systems.
\newblock {\em Computer}, 42(8):30--37.

\bibitem[Kulkarni and Rodd, 2020]{KulkarniR20}
Kulkarni, S. and Rodd, S.~F. (2020).
\newblock Context aware recommendation systems: {A} review of the state of the art techniques.
\newblock {\em Comput. Sci. Rev.}, 37:100255.

\bibitem[Liu et~al., 2016]{LiuWWLW16}
Liu, Q., Wu, S., Wang, D., Li, Z., and Wang, L. (2016).
\newblock Context-aware sequential recommendation.
\newblock In Bonchi, F., Domingo{-}Ferrer, J., Baeza{-}Yates, R., Zhou, Z., and Wu, X., editors, {\em {IEEE} 16th International Conference on Data Mining, {ICDM} 2016, December 12-15, 2016, Barcelona, Spain}, pages 1053--1058. {IEEE} Computer Society.

\bibitem[Mokhtari et~al., 2019]{DBLP:journals/access/MokhtariAZ19}
Mokhtari, G., Anvari{-}Moghaddam, A., and Zhang, Q. (2019).
\newblock A new layered architecture for future big data-driven smart homes.
\newblock {\em {IEEE} Access}, 7:19002--19012.

\bibitem[Pacheco and Hariri, 2016]{DBLP:conf/saso/PachecoH16}
Pacheco, J. and Hariri, S. (2016).
\newblock Iot security framework for smart cyber infrastructures.
\newblock In Elnikety, S., Lewis, P.~R., and M{\"{u}}ller{-}Schloer, C., editors, {\em 2016 {IEEE} 1st International Workshops on Foundations and Applications of Self* Systems (FAS*W), Augsburg, Germany, September 12-16, 2016}, pages 242--247. {IEEE}.

\bibitem[Park et~al., 2016]{DBLP:conf/milcom/ParkDISSK16}
Park, Y., Daftari, S., Inamdar, P., Salavi, S., Savanand, A., and Kim, Y. (2016).
\newblock Iotguard: Scalable and agile safeguards for internet of things.
\newblock In Brand, J., Valenti, M.~C., Akinpelu, A., Doshi, B.~T., and Gorsic, B.~L., editors, {\em 2016 {IEEE} Military Communications Conference, {MILCOM} 2016, Baltimore, MD, USA, November 1-3, 2016}, pages 61--66. {IEEE}.

\bibitem[Rafferty et~al., 2018]{DBLP:conf/iciot/RaffertyIA0HH18}
Rafferty, L., Iqbal, F., Aleem, S., Lu, Z., Huang, S., and Hung, P. C.~K. (2018).
\newblock Intelligent multi-agent collaboration model for smart home iot security.
\newblock In {\em 2018 {IEEE} International Congress on Internet of Things, {ICIOT} 2018, San Francisco, CA, USA, July 2-7, 2018}, pages 65--71. {IEEE} Computer Society.

\bibitem[Rakkappan and Rajan, 2019]{RakkappanR19}
Rakkappan, L. and Rajan, V. (2019).
\newblock Context-aware sequential recommendations withstacked recurrent neural networks.
\newblock In Liu, L., White, R.~W., Mantrach, A., Silvestri, F., McAuley, J.~J., Baeza{-}Yates, R., and Zia, L., editors, {\em The World Wide Web Conference, {WWW} 2019, San Francisco, CA, USA, May 13-17, 2019}, pages 3172--3178. {ACM}.

\bibitem[Ren et~al., 2020]{RenLLZWDW20}
Ren, R., Liu, Z., Li, Y., Zhao, W.~X., Wang, H., Ding, B., and Wen, J. (2020).
\newblock Sequential recommendation with self-attentive multi-adversarial network.
\newblock In Huang, J.~X., Chang, Y., Cheng, X., Kamps, J., Murdock, V., Wen, J., and Liu, Y., editors, {\em Proceedings of the 43rd International {ACM} {SIGIR} conference on research and development in Information Retrieval, {SIGIR} 2020, Virtual Event, China, July 25-30, 2020}, pages 89--98. {ACM}.

\bibitem[Rendle et~al., 2010]{DBLP:conf/www/RendleFS10}
Rendle, S., Freudenthaler, C., and Schmidt{-}Thieme, L. (2010).
\newblock Factorizing personalized markov chains for next-basket recommendation.
\newblock In Rappa, M., Jones, P., Freire, J., and Chakrabarti, S., editors, {\em Proceedings of the 19th International Conference on World Wide Web, {WWW} 2010, Raleigh, North Carolina, USA, April 26-30, 2010}, pages 811--820. {ACM}.

\bibitem[Rendle et~al., 2011]{RendleGFS11}
Rendle, S., Gantner, Z., Freudenthaler, C., and Schmidt{-}Thieme, L. (2011).
\newblock Fast context-aware recommendations with factorization machines.
\newblock In Ma, W., Nie, J., Baeza{-}Yates, R., Chua, T., and Croft, W.~B., editors, {\em Proceeding of the 34th International {ACM} {SIGIR} Conference on Research and Development in Information Retrieval, {SIGIR} 2011, Beijing, China, July 25-29, 2011}, pages 635--644. {ACM}.

\bibitem[Serror et~al., 2018]{DBLP:conf/IEEEares/SerrorHHSW18}
Serror, M., Henze, M., Hack, S., Schuba, M., and Wehrle, K. (2018).
\newblock Towards in-network security for smart homes.
\newblock In Doerr, S., Fischer, M., Schrittwieser, S., and Herrmann, D., editors, {\em Proceedings of the 13th International Conference on Availability, Reliability and Security, {ARES} 2018, Hamburg, Germany, August 27-30, 2018}, pages 18:1--18:8. {ACM}.

\bibitem[Sharma et~al., 2022]{DBLP:conf/nextcomp/SharmaAB22}
Sharma, M., Assotally, A., and Bekaroo, G. (2022).
\newblock Raspimonitor: {A} raspberry pi based smart home monitoring system.
\newblock In {\em 3rd International Conference on Next Generation Computing Applications, NextComp 2022, Flic-en-Flac, Mauritius, October 6-8, 2022}, pages 1--6. {IEEE}.

\bibitem[Singh et~al., 2019]{DBLP:journals/ijdsn/SinghRMKC19}
Singh, S., Ra, I., Meng, W., Kaur, M., and Cho, G.~H. (2019).
\newblock Sh-blockcc: {A} secure and efficient internet of things smart home architecture based on cloud computing and blockchain technology.
\newblock {\em Int. J. Distributed Sens. Networks}, 15(4).

\bibitem[Sun et~al., 2019]{SunLWPLOJ19}
Sun, F., Liu, J., Wu, J., Pei, C., Lin, X., Ou, W., and Jiang, P. (2019).
\newblock Bert4rec: Sequential recommendation with bidirectional encoder representations from transformer.
\newblock In Zhu, W., Tao, D., Cheng, X., Cui, P., Rundensteiner, E.~A., Carmel, D., He, Q., and Yu, J.~X., editors, {\em Proceedings of the 28th {ACM} International Conference on Information and Knowledge Management, {CIKM} 2019, Beijing, China, November 3-7, 2019}, pages 1441--1450. {ACM}.

\bibitem[Tang and Wang, 2018]{TangW18}
Tang, J. and Wang, K. (2018).
\newblock Personalized top-n sequential recommendation via convolutional sequence embedding.
\newblock In Chang, Y., Zhai, C., Liu, Y., and Maarek, Y., editors, {\em Proceedings of the Eleventh {ACM} International Conference on Web Search and Data Mining, {WSDM} 2018, Marina Del Rey, CA, USA, February 5-9, 2018}, pages 565--573. {ACM}.

\bibitem[Ur et~al., 2014]{DBLP:conf/chi/UrMHL14}
Ur, B., McManus, E., Ho, M. P.~Y., and Littman, M.~L. (2014).
\newblock Practical trigger-action programming in the smart home.
\newblock In Jones, M., Palanque, P.~A., Schmidt, A., and Grossman, T., editors, {\em {CHI} Conference on Human Factors in Computing Systems, CHI'14, Toronto, ON, Canada - April 26 - May 01, 2014}, pages 803--812. {ACM}.

\bibitem[Wang et~al., 2019]{DBLP:conf/ijcai/WangHWCSO19}
Wang, S., Hu, L., Wang, Y., Cao, L., Sheng, Q.~Z., and Orgun, M.~A. (2019).
\newblock Sequential recommender systems: Challenges, progress and prospects.
\newblock In Kraus, S., editor, {\em Proceedings of the Twenty-Eighth International Joint Conference on Artificial Intelligence, {IJCAI} 2019, Macao, China, August 10-16, 2019}, pages 6332--6338. ijcai.org.

\bibitem[Xiao et~al., 2017]{XiaoY0ZWC17}
Xiao, J., Ye, H., He, X., Zhang, H., Wu, F., and Chua, T. (2017).
\newblock Attentional factorization machines: Learning the weight of feature interactions via attention networks.
\newblock In Sierra, C., editor, {\em Proceedings of the Twenty-Sixth International Joint Conference on Artificial Intelligence, {IJCAI} 2017, Melbourne, Australia, August 19-25, 2017}, pages 3119--3125. ijcai.org.

\bibitem[Xin et~al., 2019]{XinCHWDJ19}
Xin, X., Chen, B., He, X., Wang, D., Ding, Y., and Jose, J.~M. (2019).
\newblock {CFM:} convolutional factorization machines for context-aware recommendation.
\newblock In Kraus, S., editor, {\em Proceedings of the Twenty-Eighth International Joint Conference on Artificial Intelligence, {IJCAI} 2019, Macao, China, August 10-16, 2019}, pages 3926--3932. ijcai.org.

\bibitem[Xu et~al., 2016]{DBLP:journals/cm/XuWWSM16}
Xu, K., Wang, X., Wei, W., Song, H., and Mao, B. (2016).
\newblock Toward software defined smart home.
\newblock {\em {IEEE} Commun. Mag.}, 54(5):116--122.

\bibitem[Zhan et~al., 2022]{ZhanHP022}
Zhan, Z., He, M., Pan, W., and Ming, Z. (2022).
\newblock Transrec++: Translation-based sequential recommendation with heterogeneous feedback.
\newblock {\em Frontiers Comput. Sci.}, 16(2):162615.

\bibitem[Zhang et~al., 2020]{DBLP:journals/imwut/ZhangHMZL0U20}
Zhang, L., He, W., Morkved, O., Zhao, V., Littman, M.~L., Lu, S., and Ur, B. (2020).
\newblock Trace2tap: Synthesizing trigger-action programs from traces of behavior.
\newblock {\em Proc. {ACM} Interact. Mob. Wearable Ubiquitous Technol.}, 4(3):104:1--104:26.

\end{thebibliography}



\end{document}